\documentclass[a4paper]{jpconf}	
\usepackage{graphicx}
\def\vect#1{{\mbox{\boldmath $#1$}}}

\begin{document}
\title{Octupole deformation in the nuclear chart 
based on the 3D Skyrme Hartree-Fock plus BCS model}

\author{Shuichiro Ebata$^1$ and Takashi Nakatsukasa$^2$}

\address{
$^1$Nuclear Reaction Data Centre, Faculty of Science, Hokkaido Univ., Sapporo 060-0810, Japan}
\address{
$^2$Center for Computational Sciences, Univ. of Tsukuba, Tsukuba 305-8577, Japan}

\ead{ebata@nucl.sci.hokudai.ac.jp}

\begin{abstract}
We have performed a systematic study of the ground state for even-even 
1,002 nuclei in which 28 octupole-deformed nuclei are found. 
The interplay between the spatial deformation and the pairing correlation, 
plays important roles in nuclear structure. 
Our model is based on the Skyrme Hartree-Fock plus BCS model represented in the three-dimensional 
Cartesian coordinate space which can describe any kind of nuclear shape. 
The quadrupole and octupole deformed nuclei appear in the mass region 
with characteristic neutron and proton numbers which are consistent with previous studies. 
In our results, there appear only pear shape ($\beta_{30}$) in the octupole deformed nuclei. 
We investigate the potential energy surfaces as functions of 
the octupole deformations ($\beta_{3m}$; $m=0,1,2,3$), which tells us that. 
$^{220}$Rn has also local minima in the $\beta_{31}$ and $\beta_{32}$ potential energy surfaces.
\end{abstract}

\section{Introduction}
Breaking of the rotational symmetry is spontaneously induced by the coupling between the
individual particle motions and the collective motion, which was pioneered by Bohr and Mottelson \cite{BM75}.
%A. Bohr and B. R. Mottelson, Nuclear Structure Vol.II (W. A. Benjamin, 1975). 
Furthermore, the interplay between the spatial deformation and the pairing correlation, 
which can be regarded as the deformation in the gauge space, plays important roles in nuclear structure. 
The deformation is a fundamental element to determine spectroscopic properties of nuclei. 
Systematic, non-empirical, and unrestricted studies of nuclear shape could provide 
useful insights into the nuclear structure \cite{ENI14}.
% S.Ebata, T.Nakatsukasa, and T.Inakura, Phys. Rev. C 90 024303 (2014).

Such non-empirical approaches are also important for accurate evaluation of nuclear structure data
used in various fields of nuclear application. 
In the ImPACT Program of Council for Science, Technology and Innovation, 
a project has been launched in order to investigate the nuclear reaction paths for 
the transmutation of long lived fission products (LLFP) into stable nuclei \cite{ImP}. 
% http://www.jst.go.jp/impact/en/program/08.html
Because of difficulties in experimental measurement of their reactions, 
the theoretical prediction will be necessary 
for simulation of the transmutation reactions in the bulk materials. 
For instance, the single-particle wave functions and the density
distributions obtained in the present study can be used for many kinds of
reaction studies, such as knock-out, elastic, and total reaction cross sections \cite{MWO17,HHES16}. 
%[ Minomo et al., JNST 54 (2017) 127 ].

In this work, a systematic study for the ground states of even-even 1,002 nuclei,  
is performed using the mean-field theory with a modern effective interaction. 
We use the three-dimensional (3D) coordinate-space representation with the BCS 
approximation, which can describe any shape such as non-axial octupole deformation. 
We explain briefly our models and discuss the systematic results 
for quadrupole and octupole deformations. 

\section{Model}
\subsection{Skyrme Hartree-Fock plus BCS model}
We employ the Hartree-Fock plus BCS (HF+BCS) model to investigate 
the ground state of nuclei for the whole mass region 
with the Skyrme energy density functional of SkM$^\ast$ \cite{BQB82} 
% J. Bartel, P. Quentin, M. Brack, C. Guet, and H. Hakansson, Nucl. Phys. A 386, 79 (1982).
parameter set in this work. 
The single-particle wave functions which are obtained using the imaginary-time method 
are corresponding to the canonical basis to diagonalize density matrix \cite{ENI10}. 
% S.Ebata et al., Phys. Rev. C 82 034306 (2010).
The employed pairing functional $E_{\rm pair}^\tau [\kappa^\tau, \kappa^{\tau\ast}]$ 
is the constant monopole model same as in Ref.\cite{ENI14,ENI10}. 
The $\kappa_k^\tau$ corresponds to the pair probability of each canonical-basis $\phi_k$, 
which is the product of BCS factors $u_k$ and $v_k$.  
\begin{eqnarray}
E_{\rm pair}^\tau[\kappa^\tau,\kappa^{\tau\ast}] = -\sum_{k,l>0} G_{kl}^{\tau} \kappa_k^{\tau \ast}\kappa_l^\tau,
\end{eqnarray}
where $G_{kl}^\tau$ is the pairing strength depending on the single-particle energies $\varepsilon_k, \varepsilon_l$ 
of canonical basis $\phi_k, \phi_l$; $G_{kl}=g f(\varepsilon_k)f(\varepsilon_l)$. 
The function $f$ is the cut-off function for pairing interaction 
\begin{eqnarray}
f(\varepsilon) = \left( 1 + {\rm exp}\left[ \frac{\varepsilon - \epsilon_c}{0.5} \right] \right)^{1/2} \theta(e_c - \varepsilon), 
\end{eqnarray} 
where cut-off energies $\epsilon_c = \tilde{\lambda} + 5.0$ MeV, $e_c = \epsilon_c + 2.3$ MeV, 
and $\tilde{\lambda}$ is an average energy of the highest occupied and the lowest unoccupied HF single-particle states. 
In order to determine the strength $g$, we follow a prescription of Ref.\cite{TTO96}. 
% N.Tajima, S.Takahara, and N.Onishi, Nucl. Phys. A603, 23 (1996). 
We solve the following particle-number and gap equations in each imaginary-time step, 
\begin{eqnarray}
N_\tau = \int_{-\infty}^{\infty} d\varepsilon\ \frac{(\varepsilon - \bar{\lambda}_\tau)^2 \bar{D}_\tau(\varepsilon)}{\sqrt{(\varepsilon - \bar{\lambda}_\tau)^2 + f^2(\varepsilon)\bar{\Delta}^2}}, \ \ 
\bar{\Delta} = \frac{g_\tau}{2} \bar{\Delta} \int_{-\infty}^{\infty} d\varepsilon \frac{f^2(\varepsilon) \bar{D}_\tau(\varepsilon)}
{\sqrt{(\varepsilon - \bar{\lambda}_\tau)^2 + f^2(\varepsilon)\bar{\Delta}^2}}, 
\label{eq:ND}
\end{eqnarray} 
where the value of $\bar{\Delta}$ is given by the empirical formula $12 A^{-1/2}$ MeV. 
The $\bar{D}_\tau(\varepsilon)$ is a single-particle level density in the Thomas-Fermi approximation, 
\begin{eqnarray}
\bar{D}_\tau(\varepsilon) = \frac{1}{2\pi^2} \int d\vect{r} \left( \frac{2m_\tau^\ast(\vect{r})}{\hbar^2} \right)^{3/2} 
(\varepsilon - V_\tau(\vect{r}))^{1/2} \Theta(\varepsilon - V_\tau), 
\label{eq:LD}
\end{eqnarray}
where $m_\tau^\ast(\vect{r})$ is the effective mass and $V_\tau$ means the central part of HF potential. 
Equations (\ref{eq:ND}) and (\ref{eq:LD}) fix $\bar{\lambda}_\tau$ and $g_\tau$. 

The many-body wave function is fully self-consistently obtained in the following iterative procedure: 
1) Occupied canonical states are constructed with the HF Hamiltonian $h$ according to the imaginary-time method. 
2) Calculate unoccupied canonical states inside the pairing model space using the imaginary-time method with $h$. 
3) Solve the BCS equations to obtain $u_k$ and $v_k$. 
4) Update the $h$ with new $u_k$ and $v_k$. 
5) Back to 1) and repeat the procedure until convergence. 

\subsection{Deformation parameters and constraints}
The quadrupole and octupole deformation parameters ($\beta_2, \gamma, \beta_3$) are defined as follows:
For the quadrupole deformation parameters ($\beta_2, \gamma$) are 
\begin{eqnarray}
 \beta_2 \equiv \frac{5}{4\pi A R_{\rm rms}^2}\sqrt{Q_{20}^2 + Q_{22}^2}, \hspace{3mm} 
\gamma \equiv {\rm arctan}\left[  \frac{Q_{22}}{Q_{20}} \right], \\[2mm]
Q_{20} =  \langle r^2Y_{20} \rangle,\ Q_{22} = \langle r^2 (Y_{22} + Y_{2-2})/\sqrt{2} \rangle, \nonumber
\end{eqnarray}
where $R_{\rm rms}$ is the root mean square radius. 
For the octupole deformation parameter ($\beta_3$) is 
\begin{eqnarray}
&&\beta_3 \equiv \left( \sum_{m=-3}^3 \alpha_{3m}^2 \right)^{1/2},\hspace{3mm}
\alpha_{3m} \equiv \frac{4\pi}{5}\sqrt{\frac{3A}{5}}\frac{1}{R_{\rm rms}^3} 
\int d\vect{r} r^3 X_{3m}(\Omega) \rho(\vect{r}), \\ [2mm]
&&X_{30} = Y_{30},\hspace{3mm} 
X_{3m} = (Y_{3-|m|} + Y_{3-|m|}^\ast)/\sqrt{2},\hspace{3mm} 
X_{3-m} = -i(Y_{3|m|} - Y_{3|m|}^\ast)/\sqrt{2}. 
\end{eqnarray}
The octupole component is discussed using the parameters 
$\beta_{3m} = (\alpha_{3m}^2+\alpha_{3-m}^2)^{1/2}$.
The $z$ axis is chosen so as to produce the smallest $Q_{22}$ among the three principal axes. 

To investigate the correlation in octupole deformation, 
the $r^3 X_{3m}$ are used as the constraints operator for $\beta_{3m}$.  
The constraints are added to the single-particle hamiltonian during the imaginary-time steps, 
as the quadratic form $(\langle r^3 X_{3m} \rangle - q_{3m})^2$ where $q_{3m}$ 
is a constraining expectation value corresponding to $\beta_{3m}$ \cite{RS80}.

\subsection{Numerical details}
In the 3D space, the single-particle wave function is expressed as 
$\phi (x, y, z, \sigma) = \langle x, y, z, \sigma |\phi\rangle$ with $\sigma = \pm1/2$.
The space is discretized in a cubic mesh $\Delta x = \Delta y = \Delta z$ in a sphere with a radius $R$. 
The mesh size $\Delta x$ and the radius $R$ are changed 
according to atomic number, as follows; 
for $Z = 6-18, \Delta x = 0.8$ fm with $R = 12$ fm, for $Z = 20-80$, 
$\Delta x = 1.0$ fm with $R = 15$ fm, and for $Z = 82-92$, $\Delta x = 1.0$ fm with $R = 20$ fm.
For the numerical differentiation, the nine-point formula is applied. 
The absolute binding energies may suffer from numerical error of a few MeV, 
while they provide enough accuracy for the relative energies and deformation properties.

\section{Results} 
We calculate even-even nuclei which are in the mass region with $Z=6-92$ and 
between $N=Z$ and neutron number around $N=2Z$. 
When the nucleus has a positive chemical potential in either neutron or proton, 
it is excluded from the following charts of nuclear deformation in Fig.1 and 2. 
%When the nucleus has the chemical potential less than 2.0 MeV in either neutron or proton, 
%it is excluded from subjective nuclei to avoid the gas problem. 
All the quadrupole ($\beta_2, \gamma$) and octupole deformations ($\beta_{3m}: m=0,1,2,3$) are calculated.

\subsection{Quadrupole deformed nuclei}
\begin{figure*}[h]
 \begin{center}
\includegraphics[keepaspectratio,width=8cm,angle=-90]{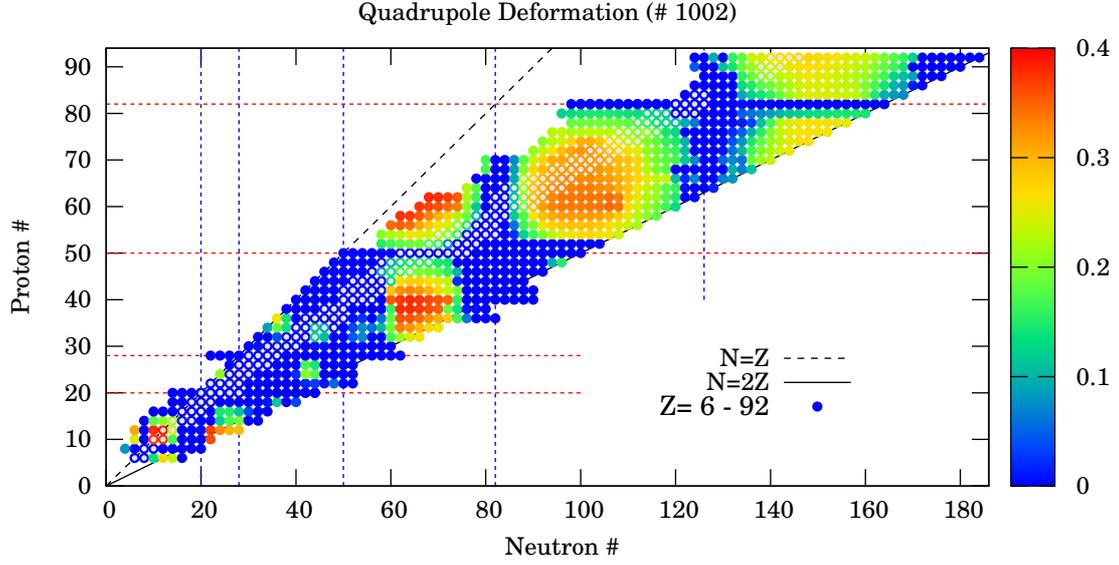}
\caption{(Color online) Absolute value of quadrupole deformation parameter $\beta_2$ 
calculated with SkM$^*$ parameter set. 
The red (blue) color means that nucleus has a well quadrupole-deformed (spherical) shape. 
See text for details.}
 \end{center}
\end{figure*}

Figure 1 shows the absolute value of quadrupole deformation parameter $\beta_2$ 
in color map for the whole nuclear chart. 
Open symbols represent stable nuclei. 
Blue symbols nuclear spherical nuclei with $\beta_2 < 0.05$. 
There is a clear correlation with spherical magic numbers. 
The deformed nuclei appear in open-shell configurations.
The distribution of quadrupole deformed nuclei is consistent with the previous systematic study \cite{MS03}, 
%M. V. Stoitsov, et. al. Phys.Rev.C 68 (2003) 054312
although there are differences in detail because the different pairing functional is used in this work.

Taking ``prolate'', ``oblate'' and ``triaxial'' nuclei as the those with $\beta_2 \geq 0.05$ with 
$\gamma < 1.5^\circ$, $58.5^\circ \leq \gamma \leq 60^\circ$ and 
$1.5^\circ \leq \gamma < 58.5^\circ$ , respectively, 54\% of even-even nuclei are quadrupole deformed. 
Among them, the ratios of prolate, oblate, and triaxial nuclei are 70\%, 12\%, and 18\%, respectively.

\subsection{Octupole deformed nuclei}
\begin{figure*}[h]
 \begin{center}
\includegraphics[keepaspectratio,width=8cm,angle=-90]{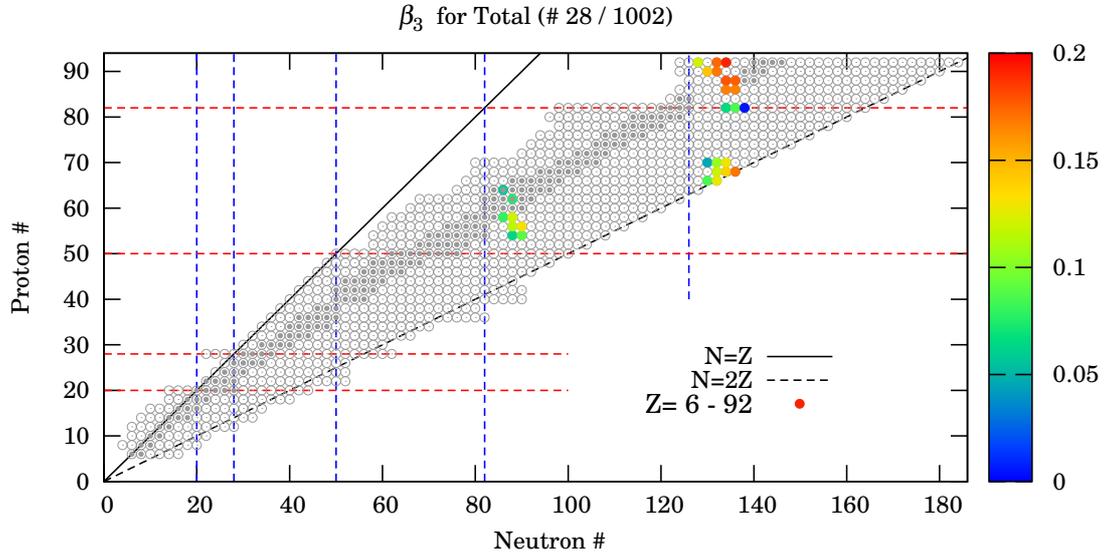}
\caption{
(Color online) Same as Fig.2, but for the absolute value of octupole deformation parameter $\beta_3$. 
}
 \end{center}
\end{figure*}
Figure 2 shows the absolute values of $\beta_3$ for nuclei with finite octupole deformation in their ground states.
%and Table 1 shows the deformation parameters ($\beta_3$, $\beta_2$, $\gamma$) of 
%octupole deformed nuclei. 
About 30 even-even nuclei appear in the nuclear chart 
which are located in the region of $84 < N < 88$ with $54 < Z < 70$, 
and $130 < N < 136$ with $84 < Z < 92$. 
These regions are consistent with those in the previous studies \cite{BN96}, 
%P.A. Butler andW. Nazarewicz, Rev. Mod. Phys. 68, 349 (1996).
and indicate that the correlation among single-particle states with $\Delta l = 3$
induces spontaneous breaking of the rotational and parity symmetries. 
Although our model allows any nuclear deformation including $\beta_{3m}: m=1,2,3$, 
we have found only the axial octupole deformation of the pear shape ($\beta_{30} > 0.000$). 

In Table 1, we show calculated deformation parameters
($\beta_3$, $\beta_2$, $\gamma$) for octupole deformed nuclei with $\beta_3\neq 0$.
They are qualitatively consistent with experimental data and
the former works, though there are some differences.
For instance,
the experimental data for $^{220}$Rn and $^{224}$Ra \cite{LPG13}
%L.P. Gaffney et al., Nature 497, 199 (2013) 
suggest that both $\beta_2$ and $\beta_3$ are larger in $^{224}$Ra
than in $^{220}$Rn, which is consistent with our results.
The deformation parameters obtained in a phenomenological analysis of 
Ref.\cite{Min06} are also consistent with our results for $^{224}$Ra. 
%N. Minkov et al., J. Phys. G 32, 497 (2006)
However, for $^{226}$Ra and $^{224,226}$Th, our calculation predicts 
the ground states with no octupole deformation.
This is different from the result of Ref.\cite{Min06}.
The number of octupole deformed nuclei in our work is smaller than the previous study 
with the macroscopic-microscopic approach \cite{Mol08}. 
This may be due to the presence of effective mass and the difference in pairing interaction. 
Actually, the number of quadrupole deformed nuclei is also smaller than Ref. \cite{MS03}. 
We understand that the pairing energy functional in the present study
is slightly stronger than the former works.
The strength of the pairing seems to be a key element for the octupole instability.
A chart of nuclear deformation of both quadrupole and octupole shapes 
will give important information for the shell structure and the pairing correlations.
A further analysis is under progress.

\begin{table}[h]
\caption{Octupole and quadrupole deformation parameters ($\beta_3$, $\beta_2$, $\gamma$) 
for octupole deformed nuclei in this work.}\ \\[-5mm]
\begin{center}
\begin{tabular}{c|ccc}
           & $\beta_3$ & $\beta_2$ & $\gamma$   \\ \hline \\[-3mm]
$^{142}$Xe &   0.061   &    0.12   &   0$^{\circ}$ \\    
$^{144}$Xe &   0.086   &    0.17   &   0$^{\circ}$ \\    
$^{144}$Ba &   0.110   &    0.16   &   0$^{\circ}$ \\    
$^{146}$Ba &   0.124   &    0.20   &   0$^{\circ}$ \\    
$^{144}$Ce &   0.076   &    0.12   &   0$^{\circ}$ \\    
$^{146}$Ce &   0.115   &    0.17   &   0$^{\circ}$ \\    
$^{150}$Sm &   0.075   &    0.20   &   0$^{\circ}$ \\    
$^{150}$Gd &   0.056   &    0.10   &   0$^{\circ}$ \\    
$^{196}$Dy &   0.078   &    0.07   &   0$^{\circ}$ \\    
$^{198}$Dy &   0.117   &    0.08   &   0$^{\circ}$ \\    
$^{200}$Er &   0.112   &    0.08   &   0$^{\circ}$ \\    
$^{202}$Er &   0.129   &    0.07   &   0$^{\circ}$ \\    
$^{204}$Er &   0.161   &    0.11   &   0$^{\circ}$ \\    
$^{200}$Yb &   0.043   &    0.05   &   0$^{\circ}$ \\ \hline
\end{tabular} \ 
\begin{tabular}{c|ccc}
           & $\beta_3$ & $\beta_2$ & $\gamma$   \\ \hline \\[-3mm]
$^{202}$Yb &   0.100   &    0.07   &   0$^{\circ}$ \\    
$^{204}$Yb &   0.120   &    0.06   &   0$^{\circ}$ \\    
$^{216}$Pb &   0.066   &    0.01   &  $-$          \\    
$^{218}$Pb &   0.088   &    0.01   &  $-$          \\    
$^{220}$Pb &   0.004   &    0.00   &  $-$          \\    
$^{220}$Rn &   0.160   &    0.13   &   0$^{\circ}$ \\    
$^{222}$Rn &   0.159   &    0.15   &   0$^{\circ}$ \\    
$^{222}$Ra &   0.170   &    0.16   &   0$^{\circ}$ \\    
$^{224}$Ra &   0.168   &    0.18   &   0$^{\circ}$ \\    
$^{220}$Th &   0.136   &    0.12   &   0$^{\circ}$ \\    
$^{222}$Th &   0.161   &    0.14   &   0$^{\circ}$ \\    
$^{220}$U  &   0.112   &    0.10   &   0$^{\circ}$ \\    
$^{224}$U  &   0.165   &    0.16   &   0$^{\circ}$ \\    
$^{226}$U  &   0.183   &    0.18   &   0$^{\circ}$ \\  \hline
\end{tabular}
\end{center}
\end{table}

\subsection{Octupole correlations}
\begin{figure*}[h]
 \begin{center}
\includegraphics[keepaspectratio,width=8cm,angle=-90]{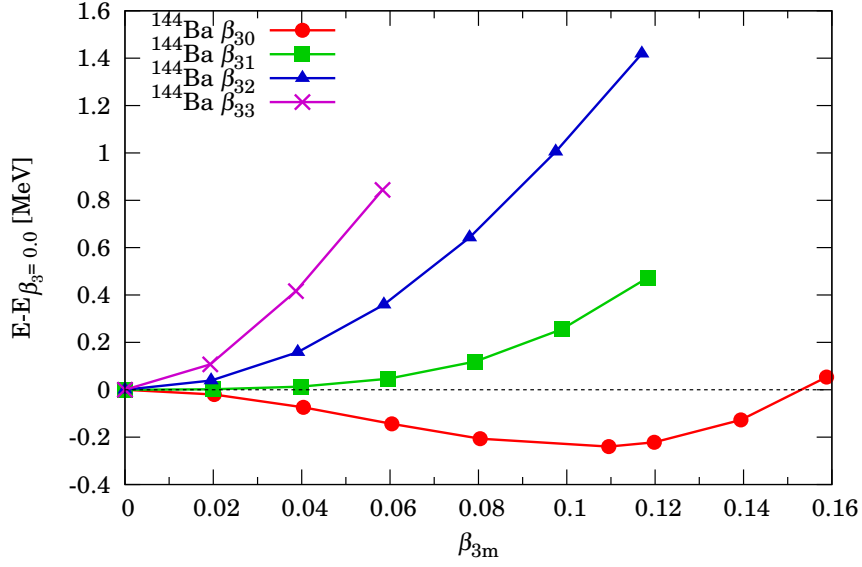}
\caption{
(Color online) Potential energy as a function of the octupole deformation parameters 
$\beta_{3m}: m=0,1,2,3$, for $^{144}$Ba. 
}
 \end{center}
\end{figure*}
To investigate the potential energy surface as a function of octupole deformation, 
the constrained HF+BCS calculations were performed. 
Figure 3 shows the potential energy surface of $^{144}$Ba 
with respect to the octupole deformation parameters. 
An experiment signature for the octupole deformation of $^{144}$Ba has been recently measured \cite{BB16}. 
%B.Bucher, et al., Phys. Rev. Lett. 116, 112503 (2016).
The circle, square, triangle and cross symbols correspond to 
the results calculated with using the octupole constraints for $\beta_{3m}: m=0,1,2,3$, respectively. 
To calculate these potential surfaces, 
we combine the constraint on $\beta_{3m} \neq 0$ together with $\beta_{3m'}=0$ $(m' \neq m)$.
The minimum at $\beta_3 \neq 0$ appears in the $\beta_{30}$ potential energy surface, 
although the correlation energy is not so large ($\sim$200 keV). 
In the other modes of octupole deformations, the lowest energy corresponds to $\beta_{3m}=0$, 
the local minimum does not appear in this work. 

\begin{figure*}[h]
 \begin{center}
\includegraphics[keepaspectratio,width=8cm,angle=-90]{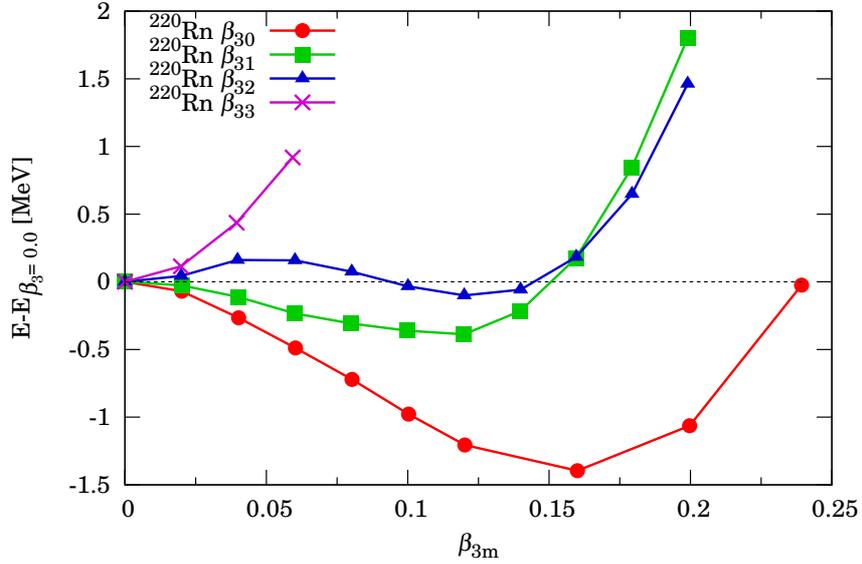}
\caption{
(Color online) Same as Fig.3, but for $^{220}$Rn. 
}
 \end{center}
\end{figure*}
Figure 4 shows the potential energy surface same as Fig.3 but for the actinide nucleus $^{220}$Rn 
which has been experimentally studied recently \cite{LPG13}. 
$^{220}$Rn also has the minimum at $\beta_{30} \neq 0$ whose correlation energy ($\sim$1.5 MeV) 
is larger than $^{144}$Ba. 
In contrast to the rare-earth nucleus $^{144}$Ba, $^{220}$Rn shows 
local minima in the potential energy curves of $\beta_{31}$ and $\beta_{32}$. 
Since the correlation energies of these local minima are relatively small (order of a few hundreds of keV), 
it is not trivial whether the state with non-axial octupole deformation exists as excited states in $^{220}$Rn. 
To answer this question, the calculation beyond the mean-field theory is necessary, 
which is beyond the scope of the present study. 

\section{Summary and Conclusion}
We investigate the ground states of even-even 1,002 nuclei using 
the self-consistent Skyrme HF+BCS model represented in the 3D coordinate space. 
The model can describe any nuclear deformation. 
The systematic investigation shows the distributions of quadrupole and octupole 
deformed nuclei in the nuclear chart. 
The location of quadrupole deformed nuclei is consistent with the previous study and 
the prolate dominance is observed at the level of 70\%. 
There are 28 nuclei which have octupole deformation in their ground state. 
The octupole deformed nuclei are located in the region with special nucleon numbers: 
$N=84 - 88, 130 - 136$ and $Z=54 - 70, 84 - 92$ which are also consistent with the previous study. 
The potential energy surface of the octupole deformed nuclei ($^{144}$Ba and $^{220}$Rn) 
with respect to the $\beta_{3m}: m=0,1,2,3$ are investigated using the constraint HF+BCS method. 
The magnitude of octupole correlation energy is much larger in $^{220}$Rn than in $^{144}$Ba.
Remarkably $^{220}$Rn has also local minima in the $\beta_{31}$ and $\beta_{32}$ potential surface. 

The octupole deformed nuclei are realized by a competing correlations 
between the shell structure and the pairing correlation. 
While the chart of the quadrupole deformation is relatively robust, 
the octupole deformation is very sensitive to details of the shell structure. 
Experimental information on the octupole correlations could be used 
to restrict the energy density functionals, 
especially on the spacing between single-particle levels with $\Delta l = 3$.

\section*{Acknowledgments}
This work was funded by ImPACT Program of Council for Science, Technology and Innovation
(Cabinet Office, Government of Japan), and was supported by Interdisciplinary Computational
Science Program in CCS, University of Tsukuba.
The computing resources for this work were supported by Research Center for Nuclear Physics, Osaka University.

\end{document}